\documentclass[aps,prl,reprint,aps,preprintnumbers,showpacs]{revtex4-2}
\usepackage{amssymb,amsmath,amsfonts,amsbsy,braket}
\usepackage{graphicx}
\usepackage{tensor}

\newcommand{\scom}[2]{\left[#1\stackrel{\star}{,}#2\right]}
\newcommand{\tr}{\mathrm{tr}}

\begin{document}

\title{Integrable spin chains in twisted maximally supersymmetric Yang-Mills theory}
\author{Tim Meier}
\email{tim.meier@usc.es}
\affiliation{Instituto Galego de F\'isica de Altas Enerx\'ias (IGFAE)
and Departamento de F\'\i sica de Part\'\i culas,\\
Universidade de  Santiago de Compostela,
15705 Santiago de Compostela,  Spain}
\author{Stijn J. van Tongeren}
\email{svantongeren@physik.hu-berlin.de}
\affiliation{Institut f\"ur Physik, Humboldt-Universit\"at zu Berlin, IRIS Geb\"aude, Zum Grossen Windkanal 2, 12489 Berlin}

\begin{abstract}
We study an angular dipole deformation of maximally supersymmetric Yang-Mills theory (SYM) that preserves its classical scale invariance. Two-point functions of suitable single trace operators, restricted to an invariant plane, are determined by scaling dimensions computable via an integrable spin chain. This spin chain is a twisted version of the famous integrable spin chain of SYM. It matches expectations from the dual string theory perfectly, presenting a precision test of holography in this new setting, and an important step to understanding general twisted integrable AdS/CFT.
\end{abstract}

%\pacs->physh things, something like: {Noncommutative field theories, Non-Abelian gauge theories, Gauge-gravity dualities} ??

\preprint{HU-EP-25/27}

\maketitle

Integrability in the AdS/CFT correspondence has given remarkable insights into both gauge and string theory, and allows high precision tests of holography \cite{Beisert:2010jr}. Starting from the discovery of an integrable spin chain in the computation of two point functions of scalar operators in planar
%maximally supersymmetric Yang-Mills theory (SYM)
SYM at one loop \cite{Minahan:2002ve}, we now have a description of scaling dimensions in planar SYM at \emph{finite} coupling via the thermodynamic Bethe ansatz \cite{Bombardelli:2009ns,Arutyunov:2009ur,Gromov:2009bc} and quantum spectral curve \cite{Levkovich-Maslyuk:2019awk} applied to the dual AdS$_5\times$S$^5$ string \cite{Arutyunov:2009ga}. These high impact results motivate a broader search for integrability, for other observables such as Wilson loops and higher point functions \cite{Bargheer:2017nne,Basso:2015zoa,Basso:2013vsa,Eden:2016xvg,Fleury:2016ykk}, and in further instances of AdS/CFT found in lower dimensions \cite{Klose:2010ki,Seibold:2024qkh} or by deformations \cite{Hoare:2021dix,Borsato:2022iip}. Here we show how a twisted integrable spin chain appears in a twist-noncommutative angular dipole deformation of planar SYM, manifesting integrability in this new setting. Its structure matches expectations from its string dual, providing important steps in extending the successful program of integrability in canonical AdS/CFT, to the broad and interesting class of its twisted deformations.

The string side of the AdS$_5$/CFT$_4$ correspondence admits a large class of integrable Yang-Baxter deformations \cite{Klimcik:2002zj,Klimcik:2008eq,Delduc:2013qra,Kawaguchi:2014qwa,vanTongeren:2015soa}, including the real-$\beta$ Lunin-Maldacena deformation \cite{Lunin:2005jy}. For homogeneous deformations, these strings are conjectured to be dual to twist-noncommutative deformations of SYM \cite{vanTongeren:2015uha,vanTongeren:2016eeb}, a large class of which has recently been explicitly constructed \cite{Meier:2023kzt,Meier:2023lku}. Twist deformations built purely on R symmetry, lead solely to the $\beta$ deformation and its three-parameter $\gamma_i$ generalization. Including spacetime and supersymmetry gives a rich landscape of new theories. It is an important open question how integrability manifests here, and how to compute observables to match their dual string counterparts. On the field theory side, twist deformations introduce star products in the SYM action, conveniently done in the formalism of \cite{Meier:2023kzt,Meier:2023lku}. We focus on a dipole analogue of the $\beta$ deformation, involving spacetime in addition to R symmetry. It preserves the classical scale invariance of SYM, leaving a clear candidate integrable spectral problem.

Our dipole product gives nonlocal angular couplings between fields, proportional to their partners' R-symmetry charges. Importantly, it leaves the spacetime-plane orthogonal to the plane of rotation involved in the deformation, untouched. This theory admits gauge invariant operators whose two-point functions -- restricted to the orthogonal invariant plane -- directly admit a twisted integrable spin chain description, a simple deformation of the one for regular SYM. For the spin chain this deformation was originally described in \cite{Beisert:2005if}, we manifest its appearance in deformed SYM. Operators at general positions are more involved, but can formally be described through our spin chain as well. Our twisted description directly matches with the dual string theory side \cite{deLeeuw:2012hp,vanTongeren:2021jhh}, providing a first nontrivial test of AdS/CFT and integrability in this novel setting.

\section{Angular dipole-deformed SYM}

Our dipole deformation arises from a noncommutative star product associated to the Drinfel'd twist
\begin{equation}
\mathcal{F}_{D} = e^{-\frac{i\lambda}{2}(R \otimes M_{23} - M_{23} \otimes R)},
\end{equation}
where $M_{23}$ is the rotation generator in the (2,3) plane, and $R$ measures the total $R$-symmetry charge. In polar coordinates in the $(2,3)$ plane, it defines the star product
\begin{equation}
\label{eq:dipolestarproduct}
\begin{aligned}
f_1(r,\theta)\star f_2(r,\theta)&\equiv \mu(\mathcal{F}^{-1} (f_1(r,\theta),f_2(r,\theta))) \\
&=e^{\frac{\lambda}{2}R_2 M_{23}}f_1(r,\theta)e^{-\frac{\lambda}{2}R_1 M_{23}}f_2(r,\theta)\\
&=f_1(r,\theta+\Lambda_2)f_2(r,\theta-\Lambda_1),
\end{aligned}
\end{equation}
where $\mu$ is the point-wise product, and $\Lambda_i=\frac{\lambda}{2}R_i$ is the (angular) dipole length of the field $f_i$, suppressing $x_0$ and $x_1$ dependence for brevity. This is the angular analogue of Cartesian dipole products \cite{Bergman:2000cw,Bergman:2001rw,Guica:2017mtd}.

We analogously deform (wedge) products of forms and spinors, twisting before multiplying conventionally \cite{Aschieri:2009zz,Meier:2023lku}. Importantly, the algebra of forms is deformed, e.g.
\begin{equation}
\phi \star\mathrm{d}x^\mu = \phi F^\mu_{~\nu}(\Lambda) \mathrm{d}x^\nu,
\end{equation}
where $\phi$ is a R-charged scalar field and $F^\mu_{~\nu}(\Lambda)$ represents a rotation in the (2,3) plane by $\Lambda$.
%The wedge product between basis forms themselves is undeformed, as they carry no R charge.

To define an action for angular dipole-deformed SYM, we introduce star products in index-free notation \cite{Meier:2023lku}
\begin{widetext}
\begin{equation}
\begin{aligned}
S^\star_{\mbox{SYM}}=&-\frac{1}{2}\tr\int \mathrm{D}\phi^{m} \wedge_\star *\mathrm{D}\phi_{m}-\frac{1}{4g_{\text{\tiny{YM}}}^2}\tr\int G\wedge_\star *G-\frac{g_{\text{\tiny{YM}}}^2}{4}\tr\int\mathrm{d}^4x~ \scom{\phi^{m}}{\phi^{n}}\star\scom{\phi_{m}}{\phi_{n}}\\
&+\tr\int\mathrm{d}^2s\mathrm{d}^2\bar{s}\int \bar{\psi}^I\star\sigma\wedge_\star*\mathrm{D}\psi_I\\
&-\frac{ig_{\text{\tiny{YM}}}}{2}~\tr\int\mathrm{d}^2s\int\mathrm{d}^4x~\sigma_m^{IJ}\psi_I\star\scom{\phi^{m}}{\psi_J}-\frac{ig_{\text{\tiny{YM}}}}{2}~\tr\int\mathrm{d}^2\bar{s}\int\mathrm{d}^4x~\sigma_{IJ}^m\bar{\psi}^I\star\scom{\phi_{m}}{\bar{\psi}^J}.
\end{aligned}
\end{equation}
\end{widetext}
This action may require modification by e.g. double trace interactions to preserve scale invariance at loop level. We assume these would not affect planar correlation functions for operators of sufficient length, by analogy to the $\beta$ and $\gamma_i$ deformation \cite{Fokken:2013mza,Fokken:2013aea,Fokken:2014soa}. This deformation preserves the part of the Noetherian symmetry of SYM that commutes with $M_{23}$ and $R$, in particular classical scale invariance. Full superconformal symmetry is realized in a twisted fashion \cite{Meier:2023lku}.

To define single-trace gauge-invariant operators
%-- analogously to other dipole theories \cite{Bergman:2000cw,Guica:2017mtd} --
we introduce the angular Wilson line $[\theta,\theta+2\Lambda]_x$ connecting $x^\mu=(x_0,x_1,r,\theta)$ and $(x_0,x_1,r,\theta+2\Lambda)$
\begin{equation}
\begin{aligned}
[\theta,\theta+2\Lambda]_x&=P\exp\left(\int_\gamma A\right)\\
\gamma^\mu(t)&=(x^0,x^1,r,\theta+2\Lambda t)^\mu,
\end{aligned}
\end{equation}
and construct single-trace gauge-invariant operators \cite{Meier:inprep,Meier:Thesis}
\begin{equation}
\label{eq:gaugeInvOperator}
\mathcal{O}_{i_1\dots i_L}(x)=\tr\left(\Phi_{i_1}\star\Phi_{i_2}\star\dots\star\Phi_{i_n}\star[\theta,\theta+2\Lambda] \right),
\end{equation}
where $\Lambda$ is the dipole length of the operator $\mathcal{O}$, $\Lambda=\sum_{i=1}^n\Lambda_i$, suppressing the uniform $x$ dependence on the right hand side. The fields $\Phi$ denote any of the (spacetime-index free) single-trace-operator building blocks of SYM: $(\phi^i,\psi_I,F)$ and their covariant derivatives. This notation skips over subtleties regarding covariance of sequences of spinors and tensors under symmetry transformations, and the definition of multiple symmetrized index-free covariant derivatives. Under the hood we use $R$ commutators and $R$ covariant derivatives \cite{Meier:inprep,Meier:Thesis} that introduce $R$-matrix entries in the component operators, and contract with basis tensors to form the above index-free single-trace operators.
%This facilitates computations, until we are ready to read off the component results.

The conjugate of \eqref{eq:gaugeInvOperator} is
\begin{equation*}
\mathcal{O}^\dagger_{i_1\dots i_L}(x)=\tr\left([\theta+\Lambda,\theta] \star \Phi^\dagger_{i_L}\star\Phi^\dagger_{i_{L-1}}\star\dots\star\Phi^\dagger_{i_1}\right).
\end{equation*}
Importantly, these gauge invariant operators are twisted cyclic \cite{Meier:inprep,Meier:Thesis} -- moving the first field to the end gives
\begin{align}
\label{eq:dipolecyclicitygeneral}
&\mathcal{O}_{i_1\dots i_L}(x)=\\
&\qquad \tr\left(e^{i \lambda (M_{23}R^{i_1}-M_{23}^{i_1}R)}\Phi_{i_2}\star\dots\star\Phi_{i_n}\star[\theta,\theta+2\Lambda]\star\Phi_{i_1}\right),\nonumber
\end{align}
where $M_{23}$ acts on all fields including the Wilson line, while $M_{23}^{i_1}$ acts on the field moved, and similarly for $R$.

\section{Two-point functions in the $(0,1)$ plane}

In undeformed SYM, two point functions are uniquely determined by their scaling dimension, and in the planar limit can be computed efficiently using integrability. Our deformation breaks four dimensional conformal symmetry, however, resulting in more complicated structure. Still, we retain conformal symmetry in the $(0,1)$ plane, where two-point functions should take the conventional form, and we might find an integrable model describing their planar scaling dimensions.

Restricting to the $(0,1)$ plane, the rotation generator in the star product only reads off the spin of the fields, and no longer touches their location. Moreover, the Wilson line in our operators trivializes, i.e.
\begin{equation}
\mathcal{O}_{i_1\dots i_L}(x^0,x^1,0,0)=\tr\left(\Phi_{i_1}\star\Phi_{i_2}\star\dots\star\Phi_{i_L}\right),
\end{equation}
and the cyclicity condition reduces to
\begin{align}
\mathcal{O}(x^0,x^1,0,0)&=\tr\left(\Phi_1\star\Phi_2\star\dots\star\Phi_n\right)\label{eq:simpledipoletwistedcyclicity}\\
&=\tr\left(e^{i\lambda\left(S_{23}R_1-R S_{23}^1\right)}  \Phi_2\star\dots\star\Phi_n\star \Phi_1\right),\nonumber
\end{align}
where $S$ now reads off the spins of the fields only \footnote{The exponential acts differently on the various components in the index free fields.}.

\begin{figure}[h]
\includegraphics[width=0.4\textwidth]{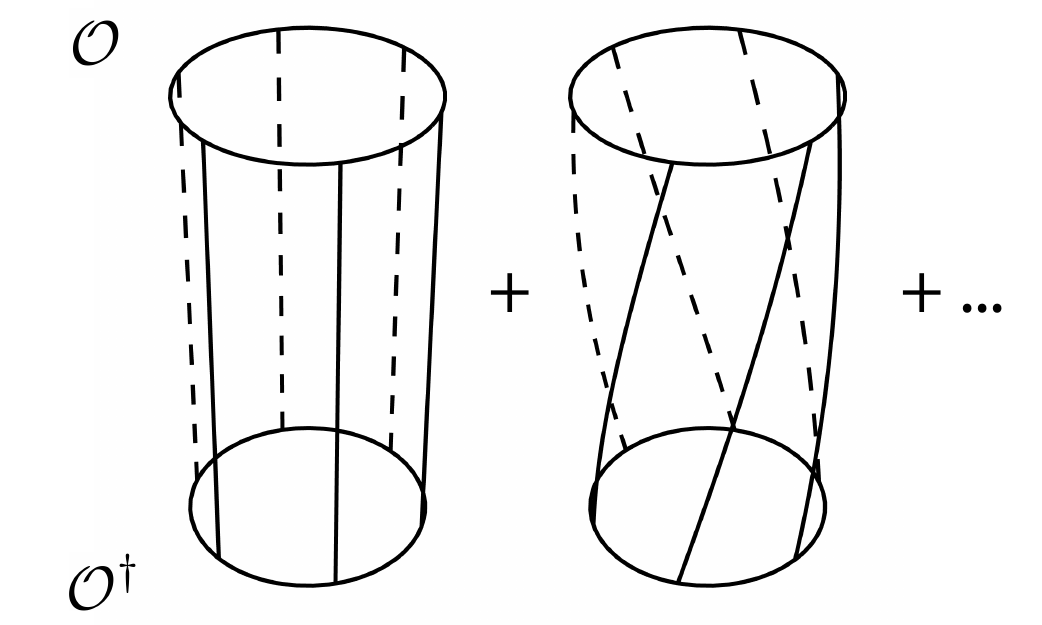}
\caption{Planar tree level contractions between two operators with six fields. Planarity allows only the canonical contraction (pictured on the left), plus cyclic shifts thereof. Each gives the same result, up to a phase, resulting in a split into twisted sectors, reflected by the sum above and in eqn. \eqref{eq:twistedsectorsplit}.}
\label{fig:twistedsectors}
\end{figure}

Now consider two-point functions in the planar limit. Contractions are as in the undeformed setting, the only difference being the presence of star products and the twisted cyclicity of operators. At tree level, planar contractions are ordered, pairwise between fields in an operator and its conjugate, plus those obtained by cyclically shifting the fields in the conjugate operator only, as in figure \ref{fig:twistedsectors}.
For the canonically ordered contraction, Lorentz invariance of the (index free) propagators allows us to cancel the star products in the operator against those in the conjugate operator with oppositely ordered product of fields \cite{Meier:2023lku}, leaving only the undeformed contraction. The twisted cyclicity introduces a nontrivial effect when the contraction is shifted, however. Rotating either operator to get a canonical and otherwise undeformed contraction, in total we get
\begin{equation}
\label{eq:twistedsectorsplit}
\begin{aligned}
\langle\mathcal{O}(x) \mathcal{O}^\dagger(y) \rangle_{\mathrm{tree}} = \frac{1}{n} &\langle\mathcal{O}(x) \mathcal{O}^\dagger(y) \rangle_{\mathrm{tree, undeformed}}\\
\times &\sum_k R^{\{1,\ldots,k\},\{k+1,\ldots,n\}},
\end{aligned}
\end{equation}
where
\begin{equation}
R^{a,b}=e^{i\lambda(R^a M^b_{23} - M^a_{23}R^b)}
\end{equation}
is the $R$ matrix evaluated on the sets of fields $a$ and $b$. For the component operators, as the result is otherwise undeformed, the different contractions each come with a distinct but simple phase factor.

Contributions to the tree level two-point function split into twisted sectors, weighed by phases from the shift to a canonical contraction. This structure persists at loop level, with one set of diagrams giving a correction to the canonically ordered tree level term, plus further cyclically shifted sets. Any such shift can again be exchanged for an overall phase by using the twisted boundary conditions, splitting into twisted sectors as at tree level. We can hence focus our attention on the untwisted sector.

\begin{figure}[h]
\includegraphics[width=0.3\textwidth]{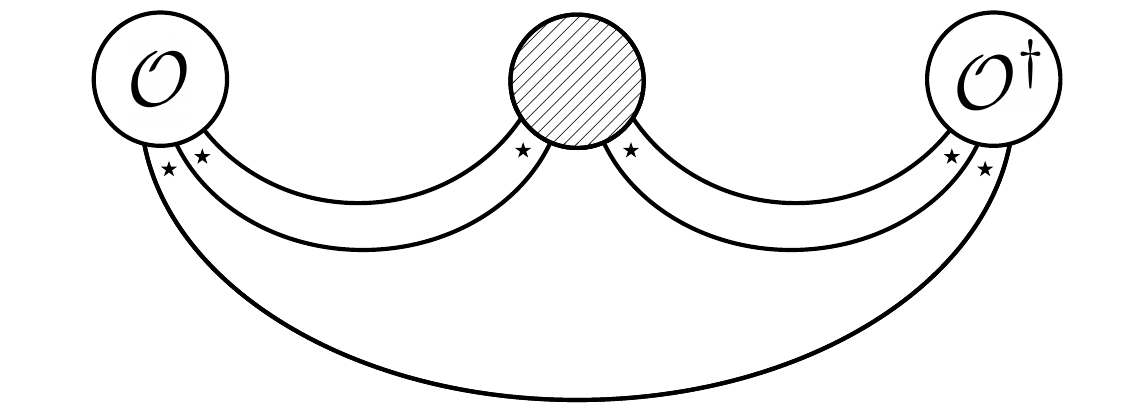}
\caption{A planar four point interaction between two operators with three fields. Star products act from left to right on subsequent lines. We removed the middle star product on the interaction term by the planar equivalance theorem. By associativity we isolate star products acting on the lines contracted with the interaction term, canceling them by Lorentz invariance of the propagators. The remaining star products cancel because the undeformed contraction via the interaction term is Lorentz invariant, like the remaining propagator.}
\label{fig:planarloopcorrection}
\end{figure}

By the planar equivalence theorem \cite{Meier:2023lku}, planar diagrams of our theory are equivalent to undeformed planar diagrams dressed with star products on external lines, and we can remove one star product freely. We then repeatedly use associativity of the star product, planarity, and Lorentz invariance of propagators and contracted undeformed diagrams, to cancel all star products, as illustrated in Figure \ref{fig:planarloopcorrection}.
%We can moreover remove one star product, keeping those between two sets of external lines only. When contracting diagrams with operators, we remove the product connecting the two sets of external lines that connect to the two operators.  Planarity and Lorentz invariance of the propagators allow us to cancel the star products in the interaction term against those between the contracted fields, using associativity of the star product to focus on these star products first (i.e. we temporarily disregard star products connecting to other fields in the operator), as indicated in Figure \ref{fig:planarloopcorrection}%\footnote{The notions of planarity related to gauge symmetry and star products are identical in this setting. We could hence also draw the usual large-N gauge theory double line diagrams, where the star products now have to align with arrows indicating index position. These star products then cancel in planar interactions, as the star products associated to any planar loop cancel, as discussed in \cite{Meier:inprep}.}
%. Any planar contraction hence gives the undeformed, \emph{Lorentz invariant} result. The remaining star products in the operators act on Lorentz invariant interaction terms or propagators, and cancel between the two operators as before.
In short, planar interactions in the untwisted sector of our theory are identical to those of undeformed SYM. The deformation appears in the twisted cyclicity of operators. Namely, for the above to apply, any interaction in the untwisted sector that crosses the end of the trace, i.e. in both operators, needs both operators to be cyclically shifted to bring it to the interior of the trace. Compared to the other undeformed interactions, these contributions pick up a phase due to the twisted cyclicity.

In spin chain language, because the bulk (trace-interior) interactions are undeformed, our twisted model is described by the integrable spin chain Hamiltonian of undeformed SYM. For (deformations of) the usual component operators, beyond the split into twisted sectors as in eqn. \eqref{eq:twistedsectorsplit}, the deformation appears solely in the twisted boundary condition that needs to be imposed on the spin chain states
\begin{equation}
\begin{aligned}
\ket{\Phi_1\Phi_2\dots\Phi_n}=e^{i\lambda\left(S_{23}R_1-R S_{23}^1\right)}\ket{\Phi_2\dots\Phi_n\Phi_1},
\end{aligned}
\end{equation}
resulting in twisted Bethe equations \cite{Beisert:2005if}. For example, two-point functions in the $\mathfrak{sl}(2)$ sector, of length $L$ with $M$ covariant derivatives, correspond to the one-loop Bethe equations
\begin{equation*}
e^{i p_k L} \prod_{j\neq k}^M \frac{u_k-u_j+i}{u_k-u_j-i} = e^{-i\lambda L}, \quad \prod_{k=1}^M e^{i p_k} = e^{-i \lambda M},
\end{equation*}
with the anomalous dimension given by the spin chain energy $E = \sum_k 1/(\tfrac{1}{4}+u_k^2)$. Our results directly build on those in undeformed planar SYM, and immediately share their reach and limitations \footnote{Up to potential subtleties for short operators due to possible double trace corrections at loop level.}. In other words, the current spin chain picture applies in the asymptotic sense of undeformed SYM.

\section{String dual and matching of spectra}

Our angular dipole deformation is conjectured to be dual to a Yang-Baxter deformed string with abelian $r$ matrix $r= M_{23} \otimes R - R \otimes M_{23}$ \cite{vanTongeren:2015uha,vanTongeren:2016eeb}. We can equivalently describe the latter as a TsT transformation of the AdS$_5$ superstring \cite{Osten:2016dvf}, similarly to the Lunin-Maldacena \cite{Lunin:2005jy} deformation.
%or Hashimoto-Itzhaki-Maldacena-Russo \cite{Hashimoto:1999ut,Maldacena:1999mh} deformations.
The resulting background mixes AdS and sphere coordinates, efficiently represented as
\begin{equation}
(g+B)_{\mu\nu} = (g_0^{-1} + \tilde{\lambda} r)^{-1}_{\mu\nu},
\end{equation}
where $g_0$ denotes the metric of undeformed AdS$_5\times$S$^5$
\begin{equation*}
\begin{aligned}
ds_0^2 = & \frac{-dt^2 + dx^2 + d\rho^2 + \rho^2 d\theta^2 + dz^2}{z^2} + (1-\mathrm{r}^2) d\phi^2\\
&  + \frac{d\mathrm{r}^2}{1-\mathrm{r}^2} + \mathrm{r}^2(d\xi^2 + \cos^2{\xi} d\chi_2^2 + \sin^2(\xi) d\chi_3^2)
\end{aligned}
\end{equation*}
and $r$ the $r$ matrix in the Killing vector representation
\begin{equation*}
r = \partial_\theta \otimes (\partial_\phi - \partial_{\chi_2} - \partial_{\chi_3}) - (\partial_\phi - \partial_{\chi_2} - \partial_{\chi_3})\otimes \partial_\theta
\end{equation*}
supported by a dilaton
\begin{equation}
e^{2(\phi-\phi_0)} = \frac{z^2}{z^2 + \tilde{\lambda}^2 \rho^2},
\end{equation}
and nontrivial RR forms. The deformation parameter $\tilde{\lambda}$ is related to the dipole one by $\lambda = \sqrt{g_{\mathrm{YM}^2}N_c} \tilde{\lambda}/2\pi$. This background has a bounded dilaton, and smoothly follows from the decoupling limit of a stack of $D3$ branes in flat space deformed by the same TsT transformation \footnote{$\mathbb{R}^{1,9}$ and AdS$_5\times$S$^5$ share the isometries used in this TsT transformation.}, while the open string picture exactly predicts our noncommutative dipole structure \cite{vanTongeren:2016eeb}. As this background preserves no supersymmetry, however, the brane configuration may be unstable, with e.g. associated tachyons in the closed string spectrum, as for the (flat space) $\gamma_i$ deformation \cite{Russo:2005yu,Spradlin:2005sv}, see also \cite{Fokken:2013aea,Bajnok:2013wsa}. For the non-supersymmetric $\gamma_i$ deformation of AdS$_5\times$S$^5$ a definite notion of duality survives in the planar limit, where it is possible to match exact spectra based on integrability \cite{Fokken:2014soa}. We expect our duality to be similarly, if not better, behaved.

The string sigma model on this background is classically integrable, with two distinct but equivalent descriptions that are well-understood in this case in particular. One uses the deformed worldsheet theory associated to the deformed background. At the quantum level this primarily manifests itself as a deformation of the factorized worldsheet S matrix of the undeformed string \cite{vanTongeren:2021jhh}. Alternatively, the deformation can be encoded in a set of twisted boundary conditions on the worldsheet \cite{Frolov:2005dj,Alday:2005ww,vanTongeren:2018vpb,Borsato:2021fuy}. This gives a quantum model described by the undeformed Bethe ansatz, with twisted boundary conditions. Both approaches match perfectly at the level of the Bethe ansatz \cite{vanTongeren:2021jhh} and string spectrum.

To compare deformed string energies to scaling dimensions in our dipole-deformed SYM, the boundary condition picture is natural. As discussed in detail in \cite{Meier:inprep}, modulo light-cone gauge fixing, the twist element that encodes the twisted boundary conditions is precisely the $R$ matrix \cite{vanTongeren:2021jhh}, matching the structure in dipole-deformed SYM. Both models hence predict matching asymptotic spectra. Moreover, it is known how to implement this twist in the thermodynamic Bethe ansatz \cite{deLeeuw:2012hp,vanTongeren:2013gva,vanTongeren:2021jhh} and quantum spectral curve \cite{Kazakov:2015efa}, providing a finite coupling connection between both models. This explicitly manifests their duality in the planar limit.

\section{Beyond the invariant plane}

Outside the $(0,1)$ plane, we lack conformal invariance, and even tree level two-point functions pick up \emph{functional} changes in twisted sectors. Moreover, when naively computing a two-point function, the derivatives that now appear in the twisted cyclicity, act on the interaction terms with unclear effect, on top of the nontrivial Wilson lines present. These complications in the direct computation, suggest that it might be more natural to use the spectral problem and spin chain in the $(0,1)$ plane to formally describe operators outside the $(0,1)$ plane.

To do so, we translate our operators to the $(0,1)$ plane, picking the origin for simplicity, and conjugate by Wilson lines to turn translation generators into covariant derivatives, which have a spin chain interpretation in the $(0,1)$ plane. I.e. we write
\begin{align}
\mathcal{O}(x)&=\tr\left((e^{x^\mu\partial_\mu}\Phi_1)\star\dots\star(e^{x^\mu\partial_\mu}\Phi_n)\star[\theta,\theta+2\Lambda]\right)\nonumber\\
&=\tr\left([x,0]\star(e^{x^\mu D_\mu}\Phi_1)\star\dots \right. \\
& \hspace{40pt} \left. \dots \star(e^{x^\mu D_\mu}\Phi_n)\star[0,x]\star[\theta,\theta+2\Lambda]\right),\nonumber
\end{align}
where $[0,x]$ denotes the straight Wilson line connecting the origin to the operator's location $x$. Moving the remaining leftmost Wilson line to the right, we find
\begin{equation}
\begin{aligned}
\mathcal{O}(x)=\tr\big((e^{x^\mu D_\mu}&\Phi_1)\star\dots\star(e^{x^\mu D_\mu}\Phi_n)\\
&\star[0,x]\star[\theta,\theta+2\lambda]\star[\tilde{x},0]\big),
\end{aligned}
\end{equation}
where $\tilde{x}=(x^0,x^1,x^2\cos(2\Lambda),x^3\sin(2\Lambda))$. The Wilson lines now combine to a Wilson loop
\begin{equation}
[0,x]\star[\theta,\theta+2\Lambda]\star[\tilde{x},0]=P\exp\left(\int_{\partial M}A \right),
\end{equation}
where $\partial M$ denotes the boundary of the disk segment surrounded by the three Wilson lines. The nonabelian Stokes' theorem \cite{Alvarez:1997ma,Broda:2000id} now gives
\begin{equation}
P\exp\left(\int_{\partial M}A \right) = \bar{P}\exp \left(\int_{M} \mathbb{G} \right) ,
\end{equation}
where $\mathbb{G}(x) = [0,x] G[x,0]$ with $G$ the field strength tensor of $A$, and $\bar{P}$ denotes a surface-ordered exponential. This Wilson loop can be expanded at the origin through $G$ and its covariant derivatives, meaning we can expand operators outside the $(0,1)$ plane, in terms of operators in this plane, with a defined spin chain interpretation. Although currently not a practical algorithm to compute two-point functions at arbitrary positions, it formally establishes a spin chain description.

%%% I WROTE FOLLOWING SOME TIME AGO, NOT SURE ANYMORE WHERE I WAS GOING WITH IT... %%%
%To find an alternative perspective on this, consider the twisted cyclicity condition \eqref{eq:dipolecyclicitygeneral}. From a hypothetical spin chain perspective, it is clear that we want to cyclically permute the fields while keeping the Wilson line in its original position. Conjugating the moved operator by the Wilson line we find
%\begin{align}
%\label{eq:dipolecyclicitygeneral}
%&\mathcal{O}_{i_1\dots i_L}(x)=\\
%&\qquad \tr\left(e^{\lambda (\hat{M}_{23}R_{i_1}-\hat{M}_{23}^{i_1}R)}\Phi_{i_2}\star\dots\star\Phi_{i_n}]\star\Phi_{i_1}\star[\varphi,\varphi+\lambda]\right),\nonumber
%\end{align}
%where the hat on $M_{23}$ indicates that it now acts via the covariant derivative in spacetime -- an object with meaning in the SYM spin chain. To deal with the

\section{Outlook}

We have demonstrated integrability in an angular dipole deformation of SYM preserving two-dimensional conformal invariance, but no supersymmetry. Two-point functions of operators restricted to the plane left invariant by the twist, take the usual massless form, allowing us to assign operators scaling dimensions. These follow from a spin chain twisted by the Drinfel'd twist defining our star product, and match perfectly with the dual deformed string energies. Operators outside the invariant plane can be expanded in terms of operators in the invariant plane, indicating a formal spin chain interpretation. These results present important steps in applying integrability in the broader landscape of twist-noncommutative SYM and their Yang-Baxter string duals, in particular giving the first full integrability account of the spectral problem of a deformation involving spacetime. 

Our results should be contrasted with \cite{Guica:2017mtd}, studying integrability for a light-like dipole deformation. The spacetime generator for this deformation is a translation generator, conceptually and technically distinct from our rotation generator. This results in a non-diagonalizable twist in the integrable model of \cite{Guica:2017mtd}, whose nontrivial structure leaves intriguing open questions regarding its exact spectrum. In contrast, our deformation corresponds to a diagonalizable twist, permitting an exact description of a full spectrum that we can perfectly match with the dual string. Our twist arises from a more involved deformation of the action and gauge invariant operators, however, building directly on recent developments.

%Our results should be contrasted with [32], studying integrability
%for a light-like Cartesian dipole deformation.
%Starting from an established action and operators, [32]
%finds a non-diagonalizable twist in the integrable model,
%whose nontrivial structure leaves several intriguing open
%questions regarding its effect on the exact spectrum. In
%contrast, we are considering a recently constructed and
%more complicated deformation of the action and operators,
%which, however, corresponds to a simpler diagonalizable
%twist, permitting an exact description of a full
%spectrum that perfectly matches the dual string.

Our deformation admits a three-parameter generalization replacing the total R charge in our twist by an arbitrary Cartan element of the $\mathfrak{su}(4)$ R symmetry. This affects the spectrum, but otherwise gives a similar model without supersymmetry. We can also replace the rotation in our twist by a boost. This introduces electric components in the $B$ field, and the brane-stack decoupling limit is no longer guaranteed to go through smoothly. We can of course still define this deformation of SYM, with a spin chain description similar to the one discussed here. We can also combine a rotation and boost, particularly a light-like combination such as $M_{01}+M_{31}$. This deformation preserves up to eight supercharges, depending on the choice of R-symmetry generator in the twist. The light-like nature of this deformation allows a nice brane construction, and our approach above readily applies, as there is still an invariant plane. However, the twisted boundary condition in the operators is now non-diagonalizable, and its effect on the integrable model and spectrum of scaling dimensions, is less clear. This model deserves further investigation, in particular in light of related recent results on the dual string side \cite{Driezen:2024mcn,Borsato:2024sru,Driezen:2025dww}.

It is our aim to use integrability to compute observables for fully spacetime-noncommutative deformations, such as the Lorentz deformation \cite{Meier:2023kzt}
%, and ultimately all deformations based on the superconformal algebra
. This deformation employs a commuting Lorentz boost and rotation, whereby the star product leaves only the origin invariant, in contrast to the invariant plane of our dipole deformation. We hence cannot have operators at distinct locations simultaneously left invariant under the star product, and consequently the conventional association of an operator at any position to a well-defined eigenstate of the dilatation operator at the origin, fundamentally breaks down. Nevertheless, one loop, planar, scalar two-point functions with one position fixed at the origin, are described by the expected undeformed spin chain \cite{Meier:inprep}, hinting at broader integrability. This analysis should be extended to general operators at generic positions. We believe there is an integrable spin chain here, although not necessarily straightforwardly in the computation of two-point functions in our basis of operators.

An integrability description of the full landscape of Yang-Baxter deformed AdS/CFT requires several developments. For deformations of SYM, the formalism of \cite{Meier:2023lku} needs to be expanded to all superconformal twists. We then need to understand which models have proper string duals, and what can be said for cases where the brane picture breaks down. To use integrability for in-depth tests of dualities, we need to understand how integrable spin chains are realized in each twisted SYM, which physical observables we can compute with them, how  we address non-diagonalizable deformations, and what the compatible integrable description is for the dual strings. On the field theory side, we might start with deformations preserving an invariant line or plane, where our approach presumably largely applies. On the string side, beyond jordanian deformations, we should systematically study abelian non-diagionalizable deformations, such as the light-cone version of our dipole deformation. Finally, from a different angle, it may be insightful to establish classical integrability of twist-noncommutative SYM in the spirit of Yangian invariance \cite{Beisert:2017pnr,Garus:2017bgl}, and study the realization of (twisted) symmetries at the quantum level.

\section{Acknowledgements.} We would like to thank Riccardo Borsato, Ben Hoare, Richard Szabo and Konstantin Zarembo for valuable discussions. The work of TM was supported by the grant RYC2021-032371-I (funded by MCIN/AEI/10.13039/501100011033 and by the European Union “NextGenerationEU”/PRTR), the grant 2023-PG083 (with reference code ED431F 2023/19 funded by Xunta de Galicia), the grant PID2023-152148NB-I00 (funded by AEI-Spain), the María de Maeztu grant CEX2023-001318-M (funded by MICIU/AEI /10.13039/501100011033), the CIGUS Network of Research Centres, and the European Union. The work of ST is supported by the German Research Foundation via the Heisenberg Programme ``Integrable Deformations in Holography'' with project number 551203197. ST is supported by LT.

\end{document}